\documentclass{article}
\usepackage{spconf,amsmath,epsfig}
\usepackage{cite}

\usepackage{graphicx}
\graphicspath{{Figures/}}
\DeclareGraphicsExtensions{.pdf,.jpeg,.png}
\usepackage{amsmath}
\usepackage{algorithm}
\usepackage{algpseudocode}

\usepackage{gensymb}
\usepackage{subfloat}

\usepackage{array}
\usepackage{multirow}
\usepackage{amsfonts}
\usepackage{stfloats}
\usepackage{url}
\usepackage[export]{adjustbox}
\usepackage{textcomp}
\usepackage{cite}

\usepackage{caption}
\usepackage[font=footnotesize]{subcaption}
\usepackage{breqn}
\usepackage{lipsum}
\usepackage{framed,multirow}
\usepackage{booktabs}
\usepackage{url}
\usepackage{xcolor}

\usepackage{comment}

\usepackage{algorithm}
\usepackage{eqparbox}
\usepackage{soul} 

\usepackage{graphicx}
\usepackage{amsmath,amssymb} 
\usepackage{bm}
\usepackage{booktabs}
\usepackage{rotating,multirow}
\usepackage{wrapfig}

\usepackage{tikz}
\usepackage{pgfplots}
\pgfplotsset{width=7cm,compat=1.16} 
\usepackage[export]{adjustbox}
\usepackage[inkscapearea=page]{svg}
\setsvg{inkscape=inkscape -z -D,svgpath=fig/}

\algnewcommand{\LeftComment}[1]{\Statex \# #1}

\makeatletter
\newcommand{\dummylabel}[2]{\def\@currentlabel{#2}\label{#1}} 
\makeatother

\graphicspath{{./Images/}{./Figures/result/}{./Images/result/}}

\usepackage{algpseudocode}

\begin{document}

    
    \title{Leveraging Membership Inference Attacks for Privacy Measurement \\ in Federated Learning for Remote Sensing Images}
    %
    \name{Anh-Kiet Duong$^{\dagger}$, Petra Gomez-Krämer$^{\dagger}$, Hoàng-Ân Lê$^ *$, Minh-Tan Pham$^ *$~\thanks{This work was done during the Master internship of Anh-Kiet Duong at IRISA, Université Bretagne Sud.}} 
    \address{
    $^{\dagger}$ L3i Laboratory, Université La Rochelle, 17042 La Rochelle Cedex 1, France\\
    $^*$IRISA, Université Bretagne Sud, UMR 6074, 56000 Vannes, France \\
    \tt\small \{anh.duong,petra.gomez\}@univ-lr.fr, \{hoang-an.le,minh-tan.pham\}@irisa.fr}
	\maketitle
	
\begin{abstract}
    Federated Learning (FL) enables collaborative model training while keeping training data localized, allowing us to preserve privacy in various domains including remote sensing. However, recent studies show that FL models may still leak sensitive information through their outputs, motivating the need for rigorous privacy evaluation. In this paper, we leverage membership inference attacks (MIA) as a quantitative privacy measurement framework for FL applied to remote sensing image classification. We evaluate multiple black-box MIA techniques, including entropy-based attacks, modified entropy attacks, and the likelihood ratio attack, across different FL algorithms and communication strategies. Experiments conducted on two public scene classification datasets demonstrate that MIA effectively reveals privacy leakage not captured by accuracy alone. Our results show that communication-efficient FL strategies reduce MIA success rates while maintaining competitive performance. These findings confirm MIA as a practical metric and highlight the importance of integrating privacy measurement into FL system design for remote sensing applications.
\end{abstract}


\section{Introduction}
\label{sec:intro}
Remote sensing imagery is a key resource for a wide range of applications, including environmental monitoring, urban planning, land-use analysis, and maritime surveillance. The increasing availability of high-resolution satellite and aerial images has driven the adoption of deep learning models to automatically extract semantic information. However, remote sensing data often contain sensitive or confidential information, making centralized data collection and model training impractical or even prohibited in many real-world scenarios.

Federated Learning (FL) \cite{zhang2021survey, chen2024federated} has emerged as a promising paradigm to address these concerns by enabling multiple clients to collaboratively train a shared model while keeping raw data localized. A general approach is to exchange model-related information instead of data samples, which preserves privacy better than centralized learning. Nevertheless, recent studies have shown that such an idea does not guarantee absolute privacy. Trained models may still leak sensitive information about their training data through inference-based attacks \cite{bai2024membership}.

Among these threats, membership inference attacks (MIAs) \cite{shokri2017membership, hu2022membership} have gained significant attention due to their effectiveness and realistic adversarial assumptions. MIAs aim to determine whether a specific data sample was included in the training dataset by analyzing the model’s output behavior. Such attacks can be conducted in a black-box setting, requiring only query access to the model, which closely reflects practical deployment conditions in FL systems. While MIAs have been extensively studied as attack mechanisms \cite{bai2024membership}, their use as a quantitative privacy evaluation metric for federated learning, particularly in remote sensing, remains underexplored. Most existing works \cite{buyuktacs2024federated, klotz2025communication, kopidaki2025federated, duong2024fedship, zhu2023privacy} focus primarily on model performance or communication efficiency, often overlooking systematic privacy assessment. 

In this work, we propose to explore MIAs as a practical and model-agnostic framework for measuring privacy leakage in FL systems.
Accordingly, we evaluate multiple black-box MIA techniques, including entropy-based attacks, modified entropy attacks, and the likelihood ratio attack (LiRA) \cite{carlini2022membership}, across different FL algorithms including FedAvg \cite{mcmahan2017communication} and FedProx \cite{li2020federated} and  features-based communication strategies \cite{duong2024leveraging} for remote sensing image classification. Through extensive experiments, we demonstrate that privacy leakage is not necessarily correlated with classification accuracy and that communication-efficient FL strategies can significantly reduce MIA success rates while maintaining competitive performance. 

In the remainder of this paper, Section \ref{sec:method} revisits the background knowledge of MIAs as well as the FL methods and communication strategies under study; Section~\ref{sec:Experiments}, then, presents our experiments and discusses the obtained results; and finally, Section \ref{sec:conclusion} concludes the paper.

\section{Methods}
\label{sec:method}

\subsection{Federated learning methods}
\label{subsec:FL}
\subsubsection{FedAvg and FedProx}
The FedAvg algorithm \cite{mcmahan2017communication} is the a simple and one of most widely used methods in FL. As presented in Algorithm \ref{alg:FedAVG}, the server receives $N$ model weights trained locally by each of the $N$ clients and aggreates them by averaging at each of the $T$ communication rounds. Each client $k$ is configured to train for the same number of $E$ local epochs, learning rate, and optimization SGD, using their own objective function $F_k(\cdot)$ before sending their weights to the server for aggregation.

\begin{algorithm}[t]
    \begin{algorithmic}[1]
	\caption{\emph{FedAvg}}
	\label{alg:FedAVG}
	\State {\bf Input:}  rounds $T$, local epochs $E$, initial weight $w^0$, number of clients $N$.
	\For  {$t=0, \cdots, T-1$}
		\State Server sends $w^t$ to all clients. 
		\State Each client trains $w^t$ on its data for $E$ local epochs, aiming to minimize objective function $F_k$ to obtain $w_k^{t+1}$.
		\State Each client $k$ sends $w_k^{t+1}$ back to the server.
		\State Server aggregates $w^{t+1} = \frac{1}{N}\sum\limits_{k=1}^{N} w_k^{t+1}$
	\EndFor
	\end{algorithmic}
\end{algorithm}

Likewise, FedProx, one of the state-of-the-art methods proposed by Li et al. \cite{li2020federated}, generalizes FedAvg to deal with data heterogeneity in a federated network \cite{zhang2021survey}. In principle, too many of local updates may cause the system to diverge due to underlying heterogeneous data. To deal with this, the authors proposed adding a proximal term $p(w)=\frac{\mu}{2}\|w-w^{t}\|^2$ to the local function $F_k(\cdot)$, to effectively constrain the impact of local updates' variability. In the special case when $\mu = 0$, FedProx becomes FedAvg. Thus, FedProx algorithm is similar to Algortihm \ref{alg:FedAVG}, except that the objective function to minimize at Step 4 is $F'_k(w) = F_k(w) + p(w)$.



\subsubsection{FL with feature-based communication}
Beyond conventional weight-sharing schemes such as FedAvg and FedProx, a recent study has proposed feature-based communication FL as an effective strategy to reduce the communication overhead while maintaining competitive performance for remote sensing image classification \cite{duong2024leveraging}. Such an approach explores alternative FL communication strategies that focus on exchanging feature representations rather than full model parameters.

In this framework, namely FedFT, each client computes average feature vectors for each semantic class from the embedding space of its local model and transmits only these compact representations to the server. The server aggregates the received class-wise features and redistributes them to participating clients. This mechanism allows clients to benefit from knowledge extracted by other participants without explicitly synchronizing network weights, significantly reducing the amount of exchanged information.


Moreover, by combining FedAvg optimization with feature communication, the classification task can be reformulated as a retrieval problem in the shared feature space, leading to improved performance while preserving communication efficiency \cite{duong2024leveraging}. Due to its reduced information exchange and reliance on high-level representations, the FedFT strategy is particularly relevant for privacy-sensitive remote sensing applications and is well suited for evaluating privacy leakage.
\subsection{Membership inference attacks}

\subsubsection{Threat model and attack scenarios}
MIAs aim to determine whether a specific data sample was included in the training dataset of a machine learning model \cite{shokri2017membership, hu2022membership}. These attacks exploit the fact that models often behave differently when predicting on training data compared to unseen samples, particularly in terms of prediction confidence. As a result, MIAs pose significant privacy risks, especially in scenarios where models are trained on sensitive data.


MIAs are commonly categorized into white-box and black-box attacks based on the adversary’s access to the target model. White-box attacks assume full access to the model architecture, parameters, and training process, enabling highly effective attacks but relying on strong adversarial assumptions. In contrast, black-box attacks only require query access to the model’s outputs, making them more realistic in practical deployments. In this work, we focus on black-box MIAs, as they reflect better real-world FL scenarios.

\subsubsection{MIA methods}
Several black-box MIA techniques have been proposed to infer membership by analyzing model outputs \cite{carlini2022membership}. In this study, we leverage three standard MIA methods including the entropy-based attack, the modified entropy attack and the likelihood ratio attack (LiRA).

The entropy-based attack leverages the observation that models typically produce lower-entropy predictions (more confident) for training samples than for non-training ones. By computing the entropy of the output probability distribution, samples with lower entropy are inferred as members of the training set. The modified entropy attack enhances the attack ability to discriminate training and non-training samples by incorporating class-dependent information and the ground-truth label. Finally, the LiRA method relies on training multiple shadow models to approximate the output distributions of models trained with and without a target sample. By fitting parametric distributions to these outputs, LiRA performs a likelihood-ratio test to infer membership. 

\subsection{Metrics in membership inference attack}
To evaluate MIA, we measure the area under the curve (AUC) of the receiver operating characteristic (ROC) curve which plots the True positive rate (TPR) against the False positive rate (FPR) at various FPR thresholds. AUC is particularly suitable for evaluating MIA \cite{carlini2022membership} since it evaluates the trade-off between the TPR and the FPR across various threshold settings. A higher AUC indicates that the model is better at distinguishing between the two classes, making it a reliable metric for assessing the effectiveness of the attack in identifying membership.

In practice, MIA is a challenging problem, resulting in generally low AUC scores for the methods involved. Consequently, relying on AUC to compare MIA techniques is not particularly effective. In addition to AUC, existing works such as \cite{carlini2022membership, chen2023overconfidence} also suggest using the \textit{TPR} at low FPR as a metric for comparing MIA methods.
\section{Experiments}\label{sec:Experiments}
\subsection{Datasets}
We evaluate our methods using two public remote sensing scene classification datasets: the UC-Merced (UCM) \cite{yang2010bag} with $21$ classes, $2100$ images ($256 \times 256$ pixels classes) and the Aerial Image Dataset (AID) \cite{xia2017aid} with $30$ classes, $10$K images ($600 \times 600$ pixels). Following \cite{duong2024leveraging}, we partition the datasets into 70/30 training and validation subsets.

\subsection{Experimental setup}
\begin{figure*}[h]
    \centering
    \includegraphics[width=0.96\linewidth]{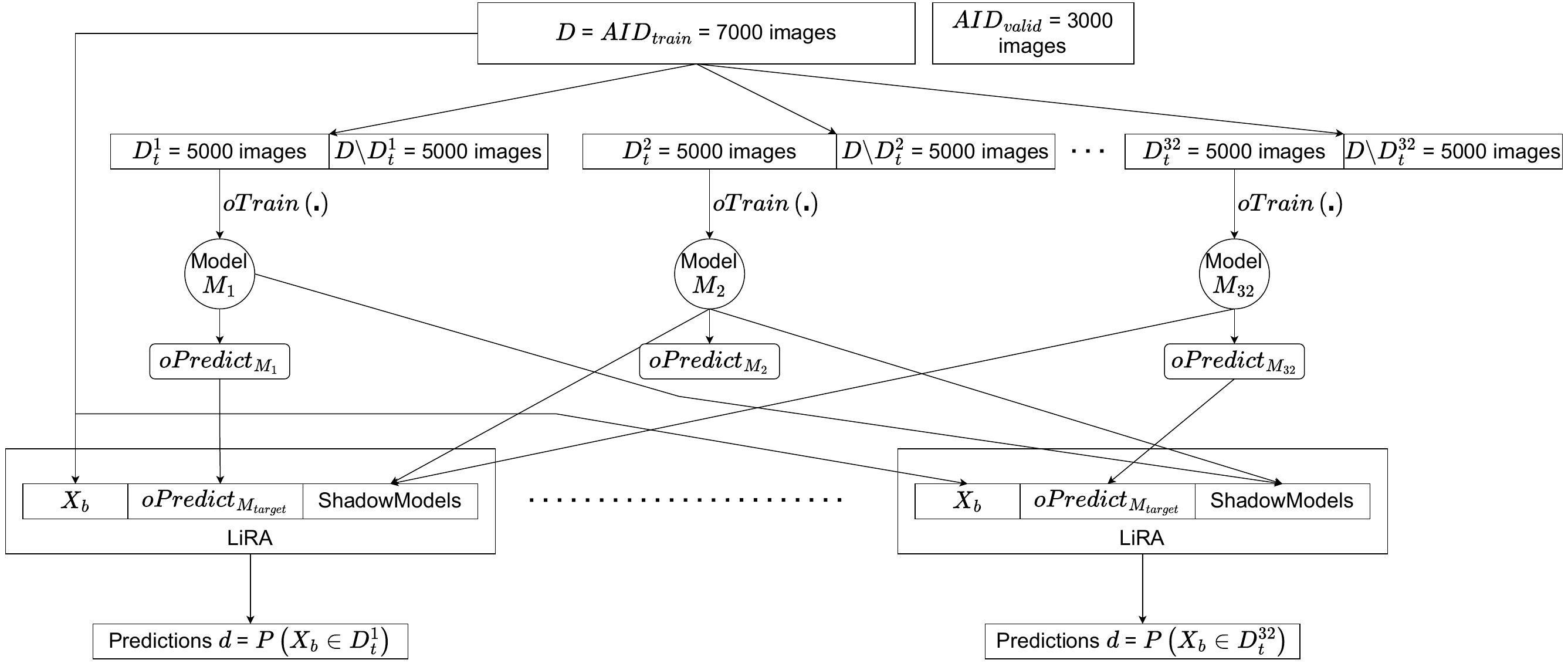}
    \caption{Illustration of our experimental setup on the AID dataset.}
    \label{fig:mia-aid}
\end{figure*}

In Figure \ref{fig:mia-aid} we illustrate how to design a cross-validation experiment on the AID dataset (a similar setup is also adopted for UCM). First, we randomly split the training subset into two parts, \textit{member} ($D_t^{i}$) and \textit{non-member} ($D \backslash D_t^{i}$), a total of 32 times. Then for each \textit{member} set $D_t^{i}$, we train a model $M_i$ by using the training function $oTrain$. Then, the prediction function $oPredict_{M_i}$ retrieves probability outputs from model $M_i$ without revealing the model weights. From these probability outputs, we apply the \textit{entropy} and \textit{modified entropy} attacks. Additionally, for each model $M_i$, we use the remaining 31 models as \textit{shadow models} for the LiRA attack. For FL we further divide the \textit{member} set $D_t^{i}$ equally among the number of clients and each client keeps 1 part.

In all our experiments, we use a consistent random seed and learning rate, and a batch size of $16$. For the classification model, the ResNet18 model is used as the network backbone with one fully connected layer added at the end to reduce the feature dimension to 128. For communication-based FL, we use the same parameter setting as done in \cite{duong2024leveraging}. 
To assess the stability of the methods, we vary the number of clients and the number of local epochs across different datasets as follows: UCM: $10$ clients with $2$ local epochs over 10 rounds; AID: $20$ clients with $4$ local epochs over 10 rounds.

We conduct experiments using the following strategies: \textit{global} centralized training; \textit{local} training where clients do not communicate with each other; two standard FL methods \textit{FedAvg} and \textit{FedProx}; and four feature communication-based FL methods \cite{duong2024leveraging} including \textit{FedFT} and \textit{FedProxFT} (sharing only features), \textit{FedFFT} and \textit{FedMFT} (for sharing both features and weights), as described in Section \ref{subsec:FL}. Except for \textit{global}, results from all other methods are averaged across clients. 

\subsection{Results}\label{sec:prires}

\begin{figure}[ht!]
  \centering
  \begin{subfigure}{0.5\textwidth}
    \centering
    \includegraphics[width=0.85\linewidth]{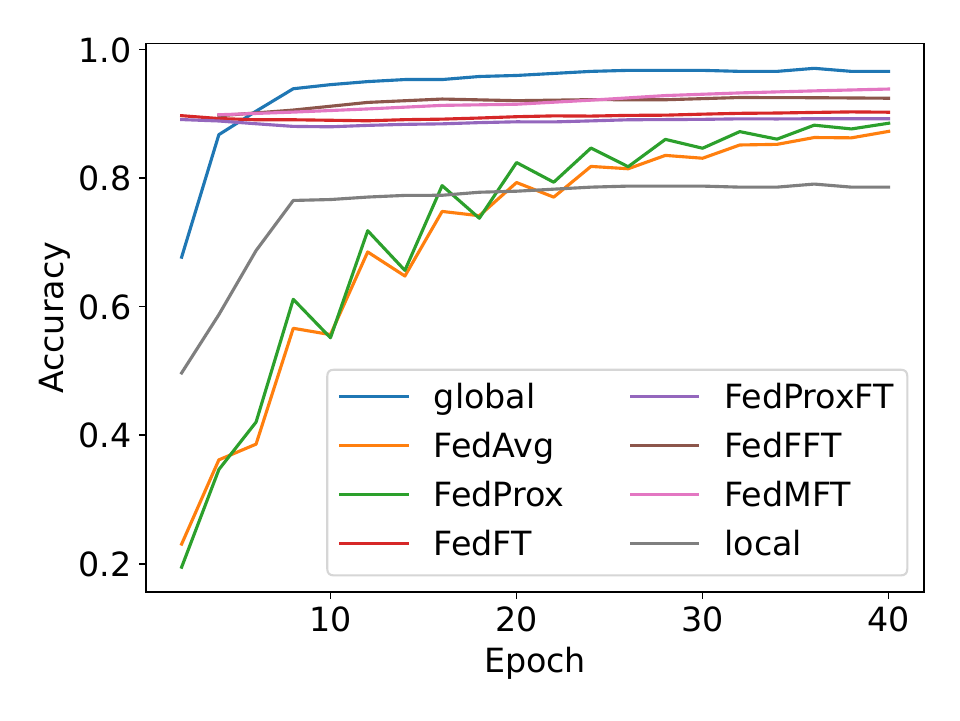}
    \caption{Performance on the UCM dataset.}
  \end{subfigure}
  \vfill
  \begin{subfigure}{0.5\textwidth}
    \centering
    \includegraphics[width=0.85\linewidth]{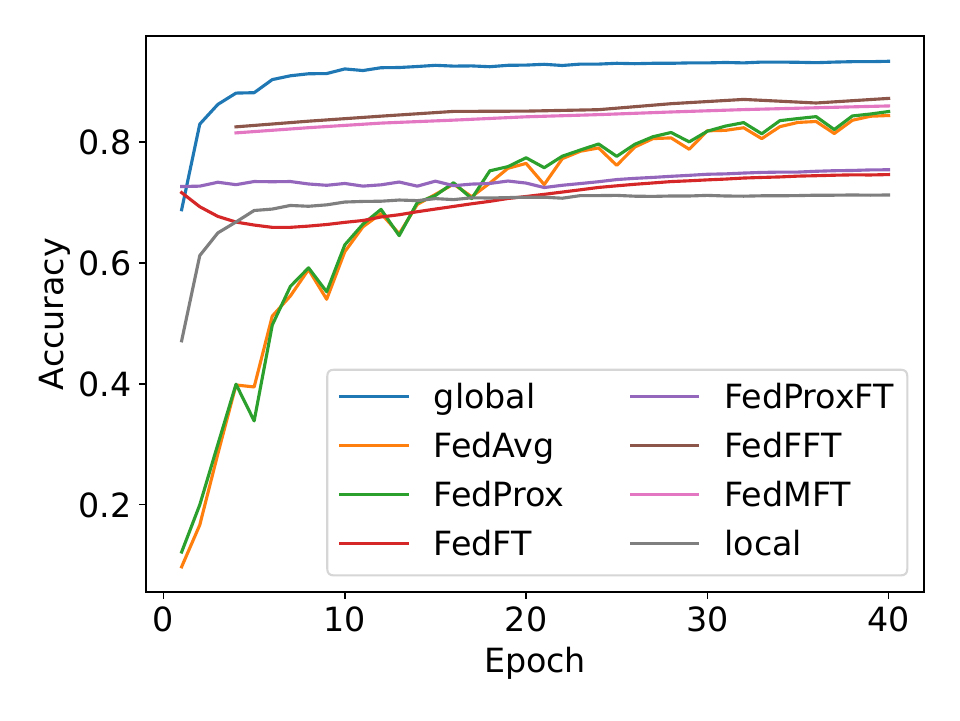}
    \caption{Performance on the AID dataset.}
  \end{subfigure}
\caption{Comparison of various training approaches.}
\label{fig:performance}
\end{figure}

Fig.~\ref{fig:performance} shows the performance of different training strategies on the two studied datasets.  Table~\ref{tab:results} presents the maximum performance among the local epochs on \textit{Accuracy} metric.

\begin{table}[h!]
\begin{subtable}{\linewidth}
\centering
\resizebox{\linewidth}{!}{\begin{tabular}{l|c|c|cccccc}
\hline
\multirow{3}{*}{Method} & \multirow{3}{*}{Accuracy} & \multirow{3}{*}{\begin{tabular}[c]{@{}c@{}}Communication\\ (Bytes)\end{tabular}} & \multicolumn{6}{c}{MIA}                                                                                    \\ \cline{4-9} 
                        &                           &                                                                                  & \multicolumn{2}{c|}{Entropy}           & \multicolumn{2}{c|}{Enropy Mod}        & \multicolumn{2}{c}{LiRA} \\ \cline{4-9} 
                        &                           &                                                                                  & AUC   & \multicolumn{1}{c|}{TPR@0.1\%} & AUC   & \multicolumn{1}{c|}{TPR@0.1\%} & AUC       & TPR@0.1\%    \\ \hline
global                  & 0.971                     & 11,560,550,400                                                                      & 0.752 & \multicolumn{1}{c|}{0.0\%}     & 0.751 & \multicolumn{1}{c|}{0.0\%}     & 0.785     & 8.71\%       \\ \hline
FedAvg                  & 0.873                     & 8,998,760,800                                                                    & 0.642 & \multicolumn{1}{c|}{0.0\%}     & 0.650 & \multicolumn{1}{c|}{0.0\%}     & 0.695     & 7.55\%       \\
FedProx                 & 0.885                     & 8,998,760,800                                                                    & 0.643 & \multicolumn{1}{c|}{0.0\%}     & 0.645 & \multicolumn{1}{c|}{0.0\%}     & 0.703     & 7.01\%       \\
FedFT                   & 0.903                     & 2,150,400                                                                        & 0.590 & \multicolumn{1}{c|}{0.0\%}     & 0.589 & \multicolumn{1}{c|}{0.0\%}     & 0.653     & 6.53\%       \\
FedProxFT               & 0.892                     & 2,150,400                                                                        & 0.589 & \multicolumn{1}{c|}{0.0\%}     & 0.589 & \multicolumn{1}{c|}{0.0\%}     & 0.641     & 6.80\%       \\
FedFFT                  & 0.925                     & 9,081,551,200                                                                    & 0.591 & \multicolumn{1}{c|}{0.0\%}     & 0.580 & \multicolumn{1}{c|}{0.0\%}     & 0.626     & 6.53\%       \\
FedMFT                  & 0.938                     & 9,009,620,320                                                                    & 0.595 & \multicolumn{1}{c|}{0.0\%}     & 0.591 & \multicolumn{1}{c|}{0.0\%}     & 0.649     & 6.60\%       \\ \hline
local                   & 0.791                     & 0                                                                                & 0.544 & \multicolumn{1}{c|}{0.0\%}     & 0.544 & \multicolumn{1}{c|}{0.0\%}     & 0.565     & 5.71\%      
\end{tabular}}
\caption{Results on the UCM dataset}
\label{tab:result UCM}
\end{subtable}
\vspace{1mm}

\begin{subtable}{\linewidth}
\resizebox{\linewidth}{!}{\begin{tabular}{l|c|c|cccccc}
\hline
\multirow{3}{*}{Method} & \multirow{3}{*}{Accuracy} & \multirow{3}{*}{\begin{tabular}[c]{@{}c@{}}Communication\\ (Bytes)\end{tabular}} & \multicolumn{6}{c}{MIA}                                                                                    \\ \cline{4-9} 
                        &                           &                                                                                  & \multicolumn{2}{c|}{Entropy}           & \multicolumn{2}{c|}{Enropy Mod}        & \multicolumn{2}{c}{LiRA} \\ \cline{4-9} 
                        &                           &                                                                                  & AUC   & \multicolumn{1}{c|}{TPR@0.1\%} & AUC   & \multicolumn{1}{c|}{TPR@0.1\%} & AUC       & TPR@0.1\%    \\ \hline
global                  & 0.934                     & 604,800,000,000                                                                  & 0.632 & \multicolumn{1}{c|}{1.8\%}     & 0.635 & \multicolumn{1}{c|}{0.0\%}     & 0.689     & 7.12\%       \\ \hline
FedAvg                  & 0.844                     & 17,999,364,800                                                                   & 0.553 & \multicolumn{1}{c|}{0.0\%}     & 0.561 & \multicolumn{1}{c|}{1.6\%}     & 0.600     & 6.30\%       \\
FedProx                 & 0.850                     & 17,999,364,800                                                                   & 0.545 & \multicolumn{1}{c|}{1.5\%}     & 0.549 & \multicolumn{1}{c|}{0.0\%}     & 0.609     & 6.16\%       \\
FedFT                   & 0.746                     & 6,144,000                                                                        & 0.524 & \multicolumn{1}{c|}{0.1\%}     & 0.526 & \multicolumn{1}{c|}{1.0\%}     & 0.580     & 5.90\%       \\
FedProxFT               & 0.754                     & 6,144,000                                                                        & 0.522 & \multicolumn{1}{c|}{1.8\%}     & 0.522 & \multicolumn{1}{c|}{1.7\%}     & 0.583     & 6.16\%       \\
FedFFT                  & 0.872                     & 18,752,004,800                                                                   & 0.568 & \multicolumn{1}{c|}{0.4\%}     & 0.551 & \multicolumn{1}{c|}{0.0\%}     & 0.597     & 6.34\%       \\
FedMFT                  & 0.860                     & 18,060,958,400                                                                   & 0.553 & \multicolumn{1}{c|}{0.0\%}     & 0.551 & \multicolumn{1}{c|}{0.0\%}     & 0.611     & 6.17\%       \\ \hline
local                   & 0.712                     & 0                                                                                & 0.518 & \multicolumn{1}{c|}{0.0\%}     & 0.528 & \multicolumn{1}{c|}{0.0\%}     & 0.542     & 5.26\%      
\end{tabular}}
\caption{Results on the AID dataset}
\label{tab:result AID}
\end{subtable}

\caption{Performance of different training strategies on the two studied datasets.}
\label{tab:results}
\end{table}

In term of classification performance, it is understandable that the \textit{global} yields the highest results while the \textit{local} results in the lowest accuracy. When comparing the FL strategies, we can categorize them into three groups with similar approach and performance: (\textit{FedFFT}, \textit{FedMFT}); (\textit{FedProx}, \textit{FedAvg}); and (\textit{FedFT}, \textit{FedProxFT}). Among these, \textit{FedFFT} and \textit{FedMFT} produce the highest results, followed by \textit{FedProx} and \textit{FedAvg}, which perform better than \textit{FedFT} and \textit{FedProxFT}. These behaviors are consistent with those in \cite{duong2024leveraging}.

In terms of communication cost, \textit{local} training is the least expensive because clients do not share information. In contrast, \textit{global} training involves clients sharing datasets, and while the experiment captures the exact bytes for each image's pixel, the actual communication cost may be reduced due to the efficiency of image compression algorithms. \textit{FedFT} and \textit{FedProxFT} transmit very few data over the network. \textit{FedMFT} is more efficient than \textit{FedFFT} because it only sends the average of features rather than all individual features. Both \textit{FedMFT} and \textit{FedFFT} have slightly higher communication cost than \textit{FedAvg} and \textit{FedProx}.

In terms of \textit{privacy}, the \textit{global} exhibits the lowest privacy, followed by the FL strategies, with \textit{local} providing the highest privacy. Among the FL strategies, \textit{FedFT} and \textit{FedProxFT} offer the best privacy, approaching that of \textit{local}, while \textit{FedProx} and \textit{FedAvg} have higher privacy than \textit{FedFFT} and \textit{FedMFT} on two out of the three datasets. It is notable that entropy-based methods have a TPR@0.1\% (TPR at $0.1\%$ FPR) that does not provide much insight. As discussed in \cite{carlini2022membership, galichin2024glira}, entropy-based methods only perform well with TPR at a high FPR. Choosing a small FPR often results in a TPR of $0.0\%$.

We can observe that as the model accuracy increases, the privacy decreases. So in the context where privacy is needed, \textit{local} is obviously the best strategy. However, if we want to cooperate in training while still achieving high privacy and low information transmission, \textit{FedProxFT} is recommended. Conversely, when performance is most important, \textit{global} is the best strategy. And in the case where we do not want to share raw data, \textit{FedFFT} is recommended.




\section{Conclusion}
\label{sec:conclusion}
This work has proposed and compared different FL approaches, including communicating only with features, combining features with pseudo-weights, and utilizing both weights and features for training and testing. Our experiments on two remote sensing  datasets showed the effectiveness of these approaches by highlighting faster convergence and reducing the amount of information transmitted over the network. This research contributes to exploring the impacts of features in various communication strategies in FL, particularly in the context of remote sensing applications.

\newpage
{\small
\bibliographystyle{ieeetr}
\bibliography{macro,ref}
}

\end{document}